\documentclass[traditabstract]{aa} 
\usepackage{txfonts}
\usepackage{graphicx}

\begin{document}

\title{Evidence for radially independent size growth of early-type galaxies in clusters
\thanks{Table 1 is only available in electronic form
at the CDS via anonymous ftp to cdsarc.u-strasbg.fr (130.79.128.5)
or via http://cdsweb.u-strasbg.fr/cgi-bin/qcat?J/A+A/}
}
\titlerunning{The radial independent size growth of early-type galaxies in clusters} 
\author{S. Andreon
}
\authorrunning{Andreon}
\institute{
INAF--Osservatorio Astronomico di Brera, via Brera 28, 20121, Milano, Italy,
\email{stefano.andreon@brera.inaf.it} 
}
\date{Accepted ... Received ...}
\abstract{
It is not well understood whether the growth of early-type cluster galaxies proceeds inside-out, outside-in, or 
at the same pace at all radii. In this work we measured the galaxy size, defined by the radius including 80\% of the galaxy light, non-parametrically. We also determined a non-parametric estimate
of galaxy light concentration, which
measures the curvature of the surface brightness profile in the galaxy outskirts.
We used an almost random sampling of a mass-limited sample formed by 128 morphologically 
early-type galaxies
in clusters with $\log M/M_{\odot} \protect\ga 10.7$ spanning the wide range $0.17<z<1.81$. 
From these data we derived 
the size-mass and concentration-mass relations, as well as their evolution.
At 80\% light radius, early-type galaxies in clusters are
about 2.7 times larger than at 50\% radius at all redshifts, and close to de
Vaucouleurs profiles in the last 10 Gyr. 
While between $z=2$ and $z=0$ both half-light and 80\% light sizes increase
by a factor of $1.7$, concentration stays constant within $2$\%, that is to say the size
growth of early-type galaxies in cluster environments proceeds at the same pace at both 
radii. 
Existing physical explanations proposed in the literature
are inconsistent with our results, demonstrating the need for dedicated numerical simulations
to identify the physical mechanism affecting the galaxy structure.
}

\maketitle

\section{Introduction}

Among massive galaxies, early-type galaxies (i.e.,  elliptical and lenticular) 
are the more abundant
population in clusters up to $z=1.2$ at least (Raichoor \& Andreon 2012).
Their size  slowly grows with time and almost doubles beginning at $z=2$ (at a fixed
$\log M/M_{\odot}=11$ mass, Andreon et al. 2016; see also Strazzullo et al. 2010)
at half the speed compared to identically selected galaxies  in the field (Andreon 2018).
It is not well understood if their size growth proceeds inside-out, outside-in, or at the same pace at
all radii.
By selecting galaxies on the red sequence, De Propris et al. (2016) find an evolution in the  
distribution of galaxy concentrations at $z>1$ due to the appearance of galaxies with very
low Sersic indices ($n\sim1$, i.e., highly concentrated). However,
the sample they studied  was not selected morphologically  and indeed some
of the high redshift concentrated galaxies are just clumpy systems (De Propris et al. 2016), which 
are rare in the local Universe. This emphasizes 
the importance of controlling for morphological composition in evolutionary studies. 
In the field environment, various works (e.g., van Dokkum et al. 2010, Patel et al. 2012) found lower Sersic indices
at high redshift and interpret this
as being due to a gradual build-up of the galaxy outer envelopes.

Major and minor dry mergers are expected to evolve galaxies in different ways in the mass-concentration plane
(Hilz et al. 2013, see also Nipoti et al. 2003); dry minor mergers mostly alter concentration,
while dry major mergers mostly change mass. 
In the IllustrisTNG simulation, galaxies with $\log M/M_{\odot}=11$ (for a Salpeter
 initial mass function (IMF)), which are mostly quiescent, have low concentrations because a large fraction of their
mass comes from stars that formed in other galaxies and preferentially deposited
in the galaxy outskirts  (Tacchella et al. 2019).
Less concentrated galaxies 
have larger fractions of stellar mass formed
ex situ, that is to say in other galaxies (Tacchella et al. 2019). 
In other terms, the concentration 
is informative about the processes that shape the structure of early-type
galaxies and
is a direct tracer of where star formation initially occurred and the
amount of mass acquired.

In this work we address the way by which galaxies grow in size 
by
measuring the radial dependence of the size growth of
early-type galaxies in clusters. We investigate the evolution of the concentration of a morphologically-selected, mass-limited sample of early-type galaxies in clusters.

Galaxies have no sharp boundaries. Traditionally, 
galaxy sizes are defined to include
an arbitrary percentage of the total flux, almost always 50\%
(see Miller et al. 2019 for an exception). This choice is arbitrary and
the inferred size evolution could be biased if galaxies
change concentration while changing size.
It is therefore important
to extend the study of the mass-size relation to other radius definitions  and
explore the possible influence of a concentration evolution. In this work we  
adopt an  80\% light size 
and we investigate the mass-size relation and its evolution with these sizes.

Two term clarifications are in order: first,
size, compactness, and concentration are different concepts; second, different
studies have different definitions of low and high concentrations.
Compactness refers to the size of
a galaxy at a fixed mass. When a galaxy is much smaller than the average for its mass,
it is called compact (or over-dense). 
Compactness is a measure of difference in the surface brightness slope compared to the average for the same mass. 
Compactness and concentration are not synonyms;
concentration is a measure of the profile curvature. Large
Sersic index $n$ are profiles for which the ratio $C_{85}=r_{80}/r_{50}$ is high, 
where $r_{x}$ is
the radius including $x$\% of the light. In other terms, large 
Sersic index $n$ profiles have extended $r_{80}$ for the same $r_{50}$ (half-light) radius.
Because of that, large Sersic indices and large $C_{85}$ are referred to in this paper as
having a low concentration. Other authors may have instead used the term high concentration for
galaxies with large Sersic indices or large $C_{85}$. Wording aside, a galaxy may be concentrated (or not) regardless of its compactness. 

Throughout this paper we assume $\Omega_M=0.3$, $\Omega_\Lambda=0.7$, 
and $H_0=70$ km s$^{-1}$ Mpc$^{-1}$. Magnitudes are in the AB system.
Results of stochastic computations are given
in the form $x\pm y$, where $x$ and $y$ are 
the posterior mean and standard deviation, respectively. The latter also
corresponds to 68\% intervals because we only summarize
posteriors close to Gaussian in this way. All logarithms are in base ten.
We use the 2003 version of Bruzual \& Charlot (2003) stellar population synthesis
models with solar metallicity and a Salpeter (1955) IMF.

\begin{figure}
\centerline{\includegraphics[width=9truecm]{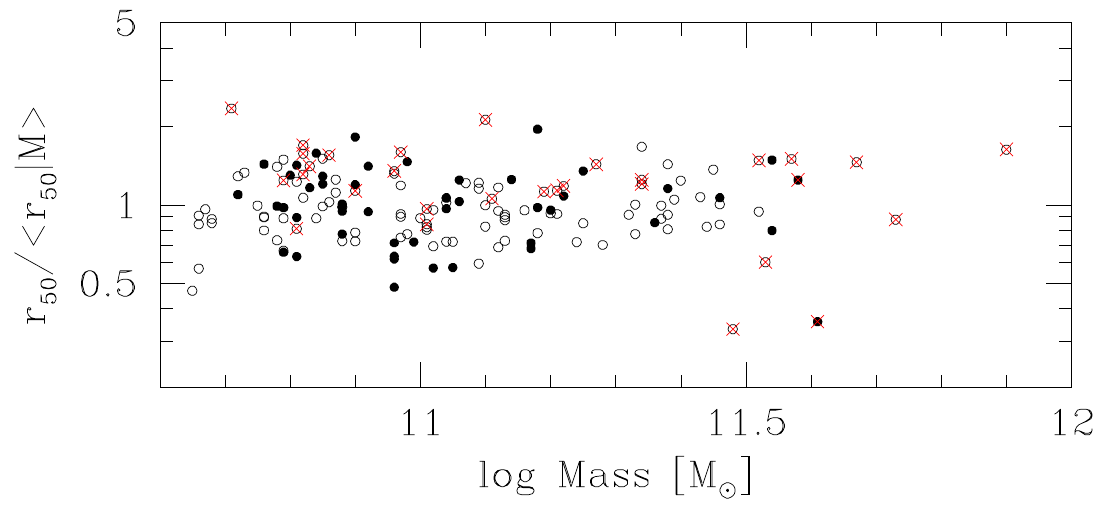}}
\caption[h]{Selection effects. 
Residuals from the mean size--mass relation vs mass. Open and closed points refer to
galaxies below and above $z=1, $ respectively.
Galaxies with missing $r_{80}$ sizes are indicated by a cross. 
}
\end{figure}

\begin{figure*}
\centerline{\includegraphics[width=9truecm]{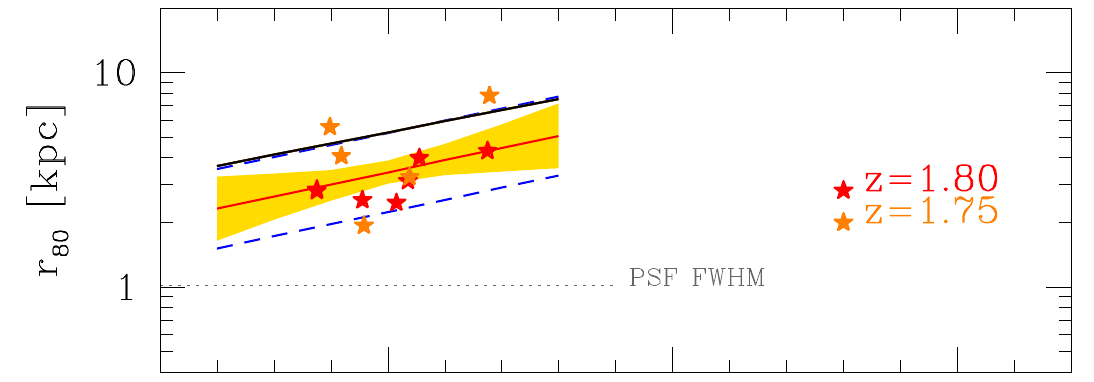}\includegraphics[width=9truecm]{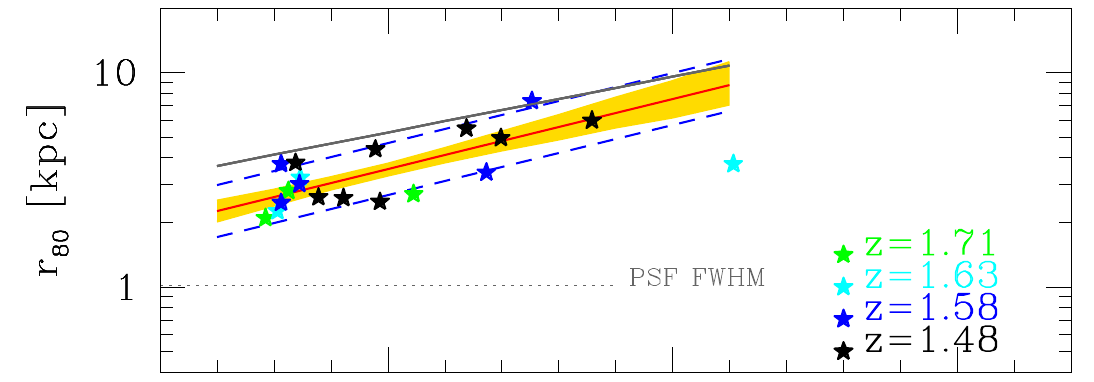}}
\centerline{\includegraphics[width=9truecm]{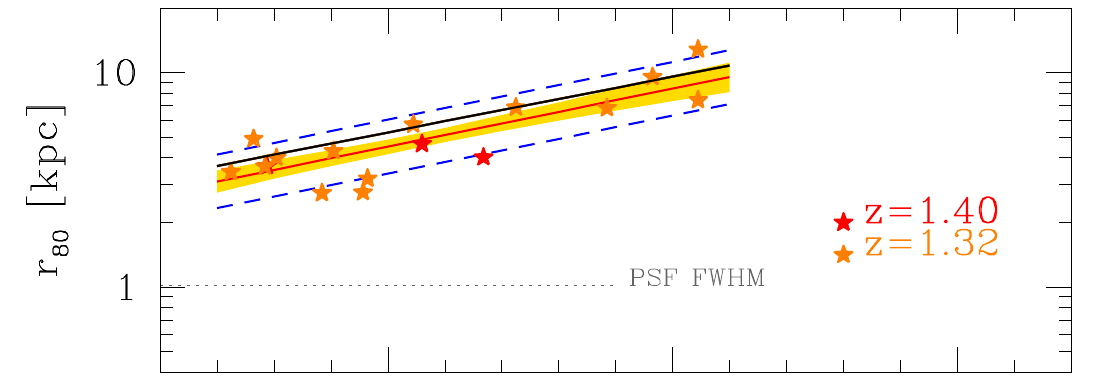}\includegraphics[width=9truecm]{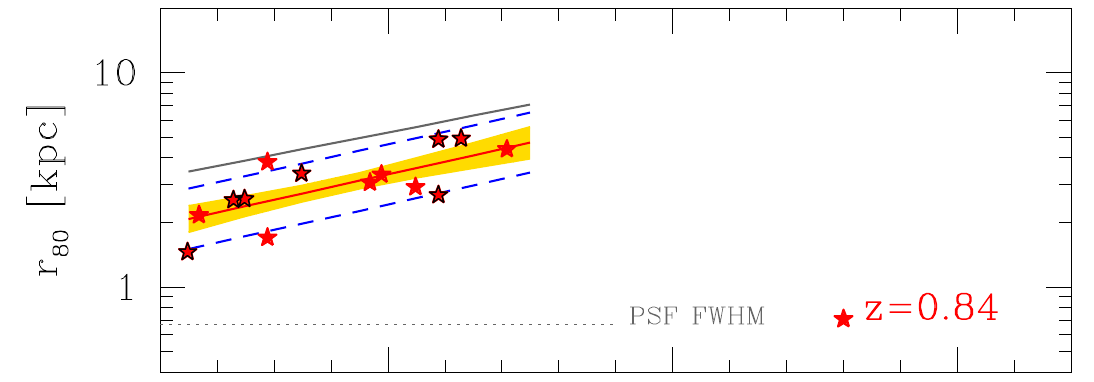}}
\centerline{\includegraphics[width=9truecm]{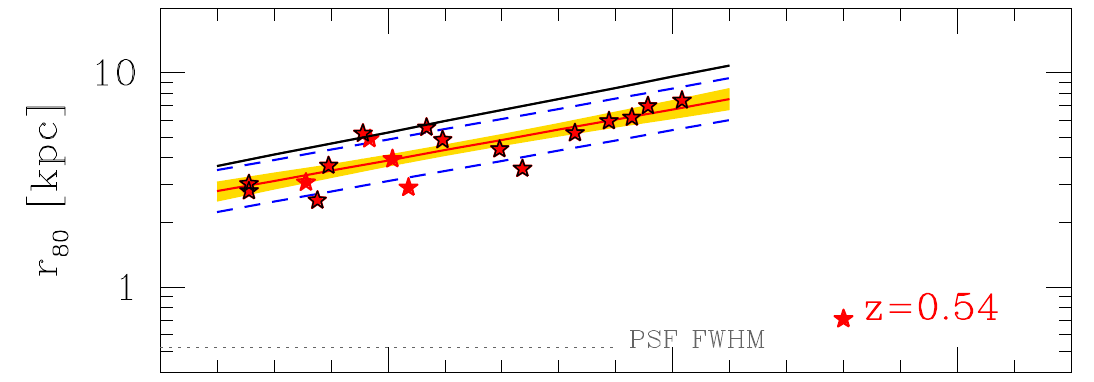}\includegraphics[width=9truecm]{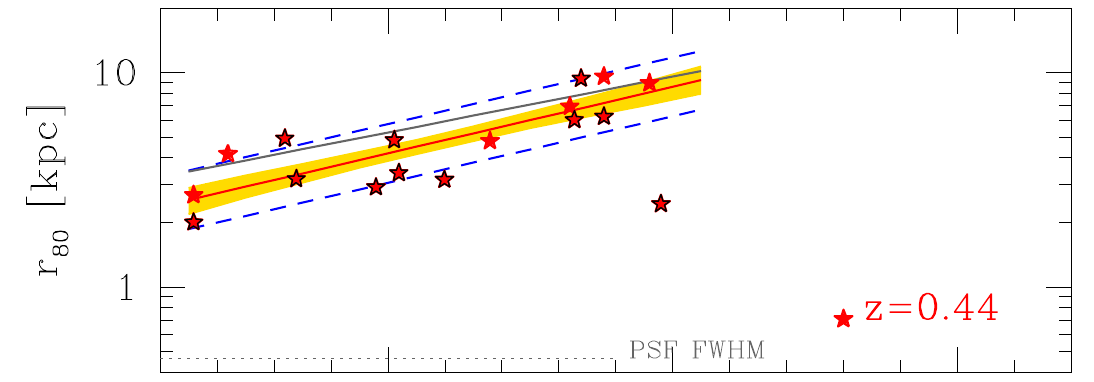}}
\centerline{\includegraphics[width=9truecm]{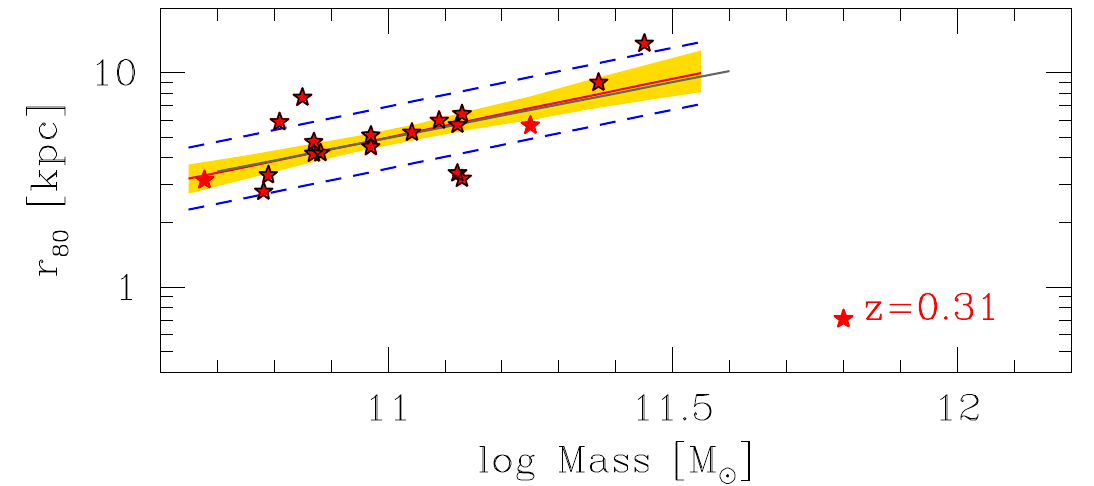}\includegraphics[width=9truecm]{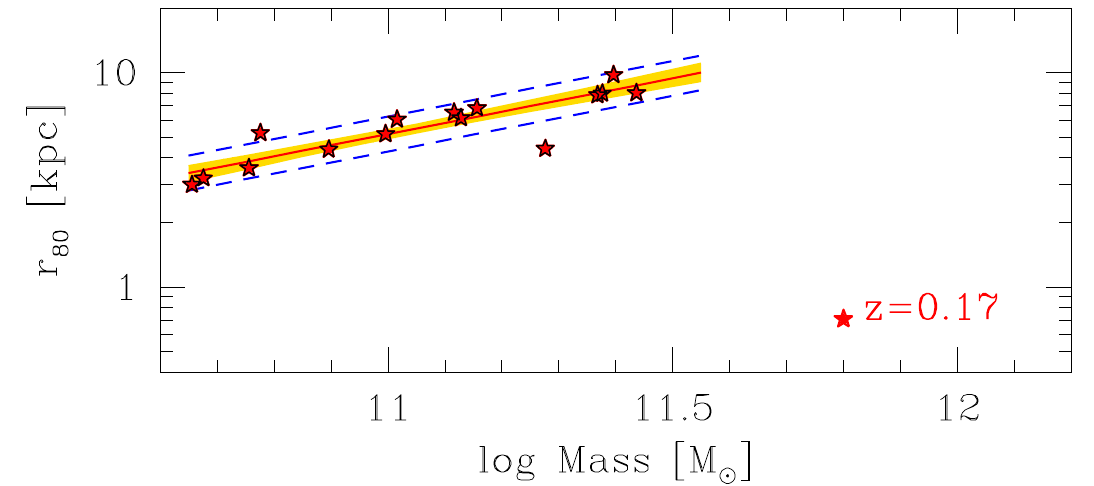}}
\caption[h]{Mass-size relation of red-sequence early-type cluster galaxies.
Points with black contours are spectroscopically confirmed galaxies.
The yellow shading indicates the 68\% uncertainty (posterior higher density) and the dashed blue 
line  indicates the corridor $\pm 1\sigma$ range around the mean model.
The solid gray line (not visible in the $z=0.17$ panel) shows the $z=0.17$ mass-size relation.
At $z<1$ points with black contours are spectroscopically confirmed galaxies.
The horizontal dotted line indicates
the PSF full width at half maximum (FWHM, below the minimal size at low redshifts). 
Sizes are corrected for (negligible) PSF blurring effects.
}
\end{figure*}

\section{Data and analysis}

\begin{center}
\begin{table}
\caption{Id, $r_{80}$, concentration, and applied PSF corrections. Coordinates,
masses, and $r_{50}$ of the same galaxies are listed in Andreon et al. (2016).}
\begin{tabular}{l r r r}
\hline
\hline
Id & $\log r_{80}$ & $\log r_{80}/r_{50}$ & PSF corr \\
  & [kpc] \\
\hline
\multispan{4}{JKCS041} \\ 
2045 & 0.45 & 0.44 & -0.02 \\ 
982 & 0.60 & 0.46 & -0.01 \\ 
988 & 0.49 & 0.35 & -0.02 \\ 
...\\
\hline             
\multispan{4}{Abell 2218 \hfill}\\ 
...\\
4286  & 0.51 & 0.40 & 0.00 \\ 
1606  & 0.48 & 0.35 & 0.00 \\ 
721 & 0.55 & 0.38 & 0.00 \\ 
\hline            
\hline
\end{tabular}                                    
\hfill \break
Table 1 is enterely available in electronic form
at the CDS via anonymous ftp to cdsarc.u-strasbg.fr (130.79.128.5)
or via http://cdsweb.u-strasbg.fr/cgi-bin/qcat?J/A+A/
\hfill \break     
\end{table}
\end{center}

\subsection{Sample selection and measurements of sizes, masses, and concentration}

The studied sample is the same one studied in Andreon, Hong \& Raichoor (2016, Paper I), 
except for a 15\% incompleteness, discussed below; 
it is formed by morphological
early-type galaxies, that is to say ellipticals and lenticulars, on the
red sequence. The sample is mass-selected, 
$\log M/M_\odot\gtrsim 10.7$ (Salpeter IMF), and  
formed by 158 galaxies at $0.17<z<1.81$\footnote{Paper I also included 
the Coma cluster. The historical data of this cluster seem to be no longer readable on
current devices, while  
more recent observations (e.g., Adami et al. 2006) are inadequate for our purposes because the background
subtraction is too aggressive,
leading to over-subtraction of the galaxy's outer regions (Andreon 2002) and an underestimate
of $r_{80}$ and of the concentration index. Therefore, the current work omitted this cluster altogether.}.
There are five clusters at $z<1$ ($z=0.175,0.306,0.439,0.54$, and $0.84$), two clusters
at $z\sim 1.35$ ($z=1.32,1.40$), four clusters at $1.5\lesssim z \lesssim 1.7$ 
($z=1.48,1.58,1.63$, and $1.71$), and two clusters
at $z\sim 1.8$ ($z=1.75,1.80$). There are between 
12 and 20 galaxies in the eight redshift bins (five at $z<1$ and three above), as detailed in Table 2. 
Clusters at close redshifts are put in a single bin, but 
our results are unaffected by redshift binning because measured quantities present, at most,
small changes with redshift and therefore the linear approximation, $\widehat{f(x)}=f(\hat{x}),$
holds (the hat indicates the mean).
At $z<1$ and at $z=1.803$ virtually all galaxies have
a spectroscopic redshift because we selected them in fields with abundant
spectroscopic coverage and we are studying massive galaxies.

In Paper I, we fit
the galaxy isophotes with ellipses plus Fourier coefficients to describe deviations
from the perfect elliptical shape in the rest-frame R-band. We classified galaxies
by detecting morphological components in the radial profiles of the isophote 
parameters. Using this morphological classification, we removed non-early-type
galaxies (spirals and irregulars) from the sample, only keeping elliptical and lenticular galaxies.
We calculated the growth curve  integrating the flux within the isophotes and extrapolated
it beyond the last measured isophote by using a library of observed growth curves. The
half-light radius $r_{50}$ is calculated as the square root of the area of the isophote, including half the flux (hence accounting
for variation in galaxy ellipticity and positional angle with radius), divided by $\pi$.
The background light, either intracluster or scattered light from bright sources, is
accounted for by fitting a low-order polynomial to the region surrounding the galaxies 
while accounting for the measuring galaxy flux at large radii. We derived the stellar mass from 
the flux in the rest-frame
R-band and an old stellar age, which is consistent with the observed color and
other works on the subject (see Paper I for details). To sample
the rest-frame R-band, we used several different filters 
across redshifts. 
An initial color selection was
adopted to save analysis time, but all galaxies with early-type morphology turned out
to be well within the selection color range, and therefore the sample is
morphologically selected. We used
HST images for all galaxies (except Coma, not used here) to achieve a resolution greater than 
1 kpc.

In the current
paper we use the radius $r_{80}$, derived again using the growth curve from the isophote including 
$\sim80$\% of the total flux ($m_t+0.25$ mag). The adopted choice of the flux percentage  
is a trade-off between  maximizing the leverage of the concentration index and minimizing the
sensitivity of large radii to errors on the background determination, or contamination
from angularly nearby objects (whose effect is negligible on the $r_{50}$, but more significant 
for lower surface brightnesses). 

Point spread function (PSF) corrections have been applied to $r_{80}$ as was already done for the half-light radius, namely
assuming an $r^{1/4}$ radial profile and convolving it with the observed PSF.
The PSF corrections are zero for virtually all galaxies
because $r_{80}$ is always much larger than the PSF.

Concentration $C_{85}=r_{80}/r_{50}$ is computed as 
the ratio between the two radii, which contain fixed fractions of the
asymptotic total galaxy luminosity and are 80\% and 50\%, respectively (e.g., Fraser 1972; de Vaucouleurs 1977). 
Galaxies with large $r_{80}$ for their $r_{50}$ have large $C_{85}$.
As a reference, a Sersic profile with $n_{Sersic}=1,4$ has $C_{85}=1.8,2.7$.
The parameter $C_{85}$ is defined regardless of the galaxy profile and irrespective
of the object profile within $r_{50}$, unlike the Sersic index; this is the reason we adopted it.

We were able to measure $r_{80}$ for $\sim 85$\% of the sample.
Figure~1 shows that the galaxies without 
a measured $r_{80}$ tend to have $\log M/M_{\odot}>11.5$. This occurs because these galaxies
are quite large and there are too many other galaxies crowding the outskirts of
the studied galaxy. This incompleteness is 
inconsequential for our analysis, which focuses on $\log M/M_{\odot}=11$ galaxies. Indeed, our analysis
de-weights the information carried by remaining galaxies having masses very different from $\log M/M_{\odot}=11$; about 15\% of the $\log M/M_{\odot}<11.45$ galaxies do not have an $r_{80}$ measurement.
The missing ones are all at $z<1$ and
tend to be larger than the average for their masses. Both effects are due to the
increased probability, with increasing size, of having several overlapping galaxies.
The crowding makes isophotal analysis at $r_{80}$ unfeasible. The redshift dependence occurs 
because at a given mass galaxies are intrinsically smaller at high redshift.
Since missing galaxies are at $z<1$, and galaxies with larger $r_{80}$ for their $r_{50}$ are
likely more affected by interlopers, (selective) incompleteness induces a redshift-dependent bias
on concentration that we need to account for. To estimate it, we computed how biased the
estimates of location and scatter of a distribution systematically missing 15\% are. More precisely, we drew values from a zero-mean Student-t distribution (used for real data 
to model the scatter) and we randomly removed 15\% of those on 
the positive side to mimic the fact that missing galaxies have larger-than-average sizes. 
With such a biased sample, the derived mean is biased low by 8\%, while the scatter is biased
by less than 1\%. The former is comparable to our errors, and therefore applied to our
results at $z<1$; the latter is negligible and was therefore disregarded. We note that our small bias correction
actually slightly overestimates the effect of incompleteness because the latter is
not as sharp as the step function adopted in the simulation.

\section{Results}

\subsection{$r_{80}$-mass scaling}

Table 1 lists id, $r_{80}$, concentration $C_{85}$, and applied  
PSF correction to $r_{80}$ of the 128 early-type galaxies studied in this work. Coordinates, masses,
and half-light radii are listed in Paper I. Figure~2 shows the $r_{80}$-mass relation at various redshifts.

The scatter at a given mass
is clearly non-Gaussian, at least because of the presence of
few outliers (e.g., the most massive galaxy of the cluster 
at $z=0.44$ in Fig.~2). Therefore,
we fit the mass-size relation modeling the scatter
around it with a Student-t distribution with ten degrees of freedom to limit the
impact of outliers, as already done in Andreon (2012) to model the metallicity 
scatter. In the fit, we also leave the slope free in order to
allow different evolutions at different masses and to de-weight $\log M/M_\odot\gg 11$ 
galaxies, because we want to focus on 
$\log M/M_\odot\sim 11$ galaxies (see Paper I for technical details). Fit results are listed in
Table~2 and displayed in Fig.~2. 
Figure 2 also shows the $z=0.17$ mass-size relation as solid gray line.
As for the more studied $r_{50}$ sizes, galaxies seem to become smaller
with increasing redshift.

The mean size at $\log M/M_\odot=11$ after (minor) selection-effect corrections is shown in Fig.~3
as a function of redshift. To quantify the possible decrease in mean size with
redshift, we fit a linear relation with uniform priors
on intercept and angle. We find:

\begin{equation}
\log r_{80|\log M = 11} = 0.70\pm0.01 -(0.11\pm0.02) (z-0.25)
\end{equation}
as depicted in Fig.~3. The found redshift dependence, $-0.11\pm0.02,$ is less than $1\sigma$
away from that found for the half-light radius, $-0.13\pm0.02$ in Andreon (2018), also depicted in
Fig.~3; this indirectly suggests a minor, at most, evolution in concentration.

Figure~4 compares the scatter around the mass-size relation when using both $r_{50}$ and
$r_{80}$. 
There is a weak indication\footnote{The two-side p-value
is 0.01 to 0.07 according to binomial statistics, depending on whether or not the
ex-equo at $z=0.55$ is considered.} for a tighter
mass-size relation when using $r_{80}$ compared to when using $r_{50}$, but
more data are needed to strengthen the 
evidence\footnote{We updated
the derivation of the scatter of the $r_{50}$-mass relation in Paper I
to make it consistent with the present work.}.

\begin{figure}
\centerline{\includegraphics[width=9truecm]{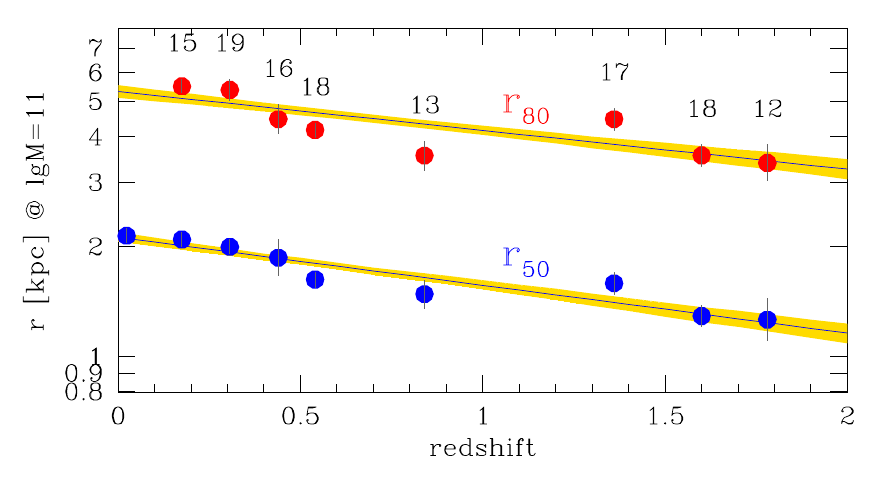}}
\caption[h]{Sizes at $\log M/M_\odot=11$ vs redshift. Red and blue points are $r_{80}$and $r_{50}$
sizes, respectively. The number above the
points indicates the number of galaxies. 
The solid line and shading show the fitted 
relation and its 68\% uncertainty (posterior highest density interval).
}
\end{figure}

\begin{figure}
\centerline{\includegraphics[width=9truecm]{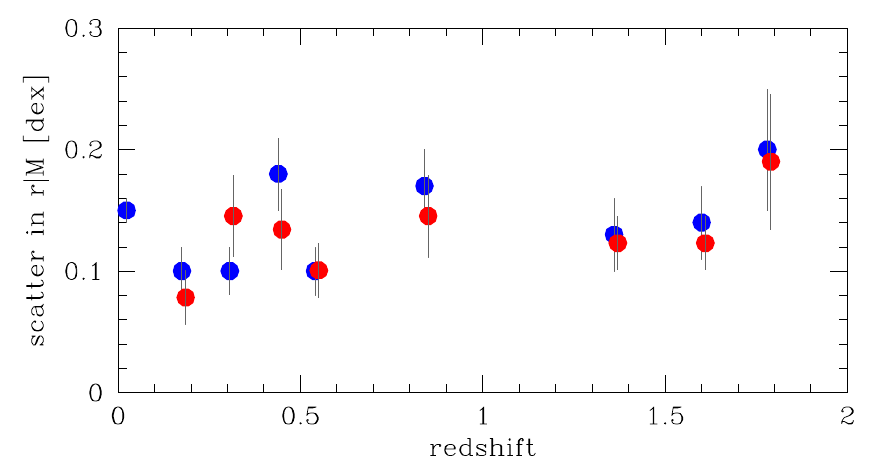}}
\caption[h]{Scatter around the size-mass relation vs redshift. Red and blue points refer to $r_{80}$ and $r_{50}$
sizes, respectively.  Scatter is computed from Student-t scale $s$ as $\sigma = s *\sqrt{10/8}$.
}
\end{figure}

\subsection{Concentration-mass scaling}

Figure~5 shows the $C_{85}$-mass relation of red-sequence early-type galaxies.
As for the $r_{80}$-mass relation,
we adopted a Student-t
distribution with ten degrees of freedom to model the scatter around the mean relation.  
In the fit, we also left the slope free so as to not force 
the same evolution at different masses and to de-weight $\log M/M_\odot\gg 11$ 
galaxies. One extreme outlier, the most massive galaxy in the $z=1.60$ sample, 
is flagged by hand because, lying at masses unsampled by other
data, it severely affects the derived slope. 
Fit results are listed in
Table~2 and also displayed in Fig.~5. 
The mean concentration index
at $\log M/M_\odot=11$ is shown in Fig.~6
as a function of the redshift, and fitted as was done for the mean $r_{80}-z$ relation. 
We find:

\begin{equation}
\log C_{85|\log M = 11} = 0.446\pm0.007 +(0.001\pm0.008) (z-0.25)
\end{equation}
as depicted in Fig.~6. At a fixed mass, early-type galaxies on the red sequence 
have not changed concentration in the the last 10 Gyr.

The right ordinates of Figs.~5 and 6 show approximated Sersic indices corresponding to the
measured concentration index. These are derived from the growth curve of 
simulated galaxies with Sersic profiles for illustrative purposes only. In the presence of a background,
the determination of $C_{85}$ and Sersic index $n_{sersic}$ depends on the precise way the background
is estimated for galaxies with large values of $C_{85}$ or Sersic index. 
For example, estimating the background
at three times the effective radius leads to a strongly biased determination of concentration
and Sersic index of $n_{ser}\ga 4$ (because $r_{80} \approx 3 r_{50}$). 
Our $C_{85}$-$n_{sersic}$ conversion assumes Sersic
profiles and that the background
is measured at $\approx 7 r_{50}$. We always use $C_{85}$ in our analyses and the 
conversion to the Sersic index is only shown in
our figures for a qualitative appreciation
of the sensitivity of our measurement on a scale commonly used.  

Figure~7 shows the scatter in concentration as a function of redshift. The scatter is about 15\% and
quite constant with redshift, indicating that the population spread
is largely time-independent.

\begin{center}
\begin{table}
\caption{Results of the various $y=\beta x + \alpha$ fits with scatter given
by a Student-t distribution with scale $s$, before corrections for selection effects.}
\begin{tabular}{l l l r l l l l}
\hline
 z    & $\alpha$ & err  & $\beta$ &  err & $s$ & err & $N_{gal}$ \\
\hline
\multispan{8}{\hfill $\log r_{80}$ vs $(\log M -11)$ \hfill}\\
0.175  & 0.71 & 0.02&  0.52 & 0.08 & 0.07 & 0.02  &    15     \\
0.306  & 0.70 & 0.03&  0.55 & 0.16 & 0.13 & 0.03  &    19     \\
0.439  & 0.62 & 0.04&  0.62 & 0.12 & 0.12 & 0.03  &  16       \\
0.54   & 0.59 & 0.02&  0.48 & 0.09 & 0.09 & 0.02  &    18     \\
0.84   & 0.52 & 0.04&  0.59 & 0.20 & 0.13 & 0.03  &  13       \\
1.36   & 0.65 & 0.03&  0.54 & 0.11 & 0.11 & 0.02  &  17       \\
1.60   & 0.55 & 0.03&  0.64 & 0.15 & 0.11 & 0.02  &    18     \\
1.78   & 0.53 & 0.05&  0.56 & 0.48 & 0.17 & 0.05  &    12     \\
\hline
\multispan{8}{\hfill $\log (r_{80}/r_{50})$ vs $(\log M -11)$ \hfill}\\
0.175  & 0.42 & 0.01 & 0.09  &  0.05 & 0.05 & 0.01 &   15 \\
0.306  & 0.40 & 0.02 & 0.02 &  0.08 & 0.07 & 0.01 &   19 \\
0.439  & 0.41 & 0.02 &  0.07 &  0.07 & 0.07 & 0.01 &   16 \\
0.54   & 0.41 & 0.02 & -0.09 &  0.06 & 0.06 & 0.01 &   18 \\
0.84   & 0.41 & 0.02 & -0.17 &  0.11 & 0.07 & 0.02 &   13 \\
1.36   & 0.46 & 0.01 & -0.02 &  0.05 & 0.06 & 0.01 &   17 \\
1.60   & 0.44 & 0.01 & 0.00  &  0.07 & 0.04 & 0.01 &   17 \\
1.78   & 0.44 & 0.02 & -0.08 &  0.17 & 0.06 & 0.02 &   12 \\
\hline                                                    
\hline
\end{tabular}                                    
\hfill \break
The standard deviation $\sigma$ of a Student-t distribution with
10 degree of freedoms and scale $s$ is given by $\sigma = s *\sqrt{10/8}$
\hfill \break     
\end{table}
\end{center}

\begin{figure*}
\centerline{\includegraphics[width=9truecm]{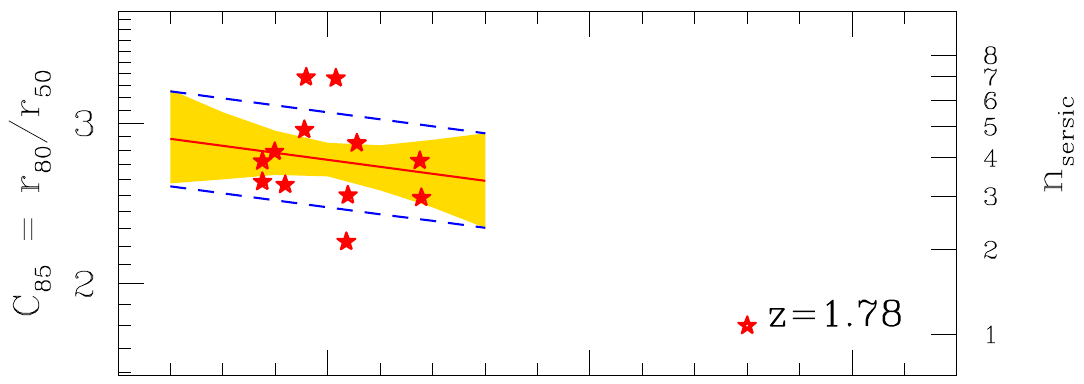}\includegraphics[width=9truecm]{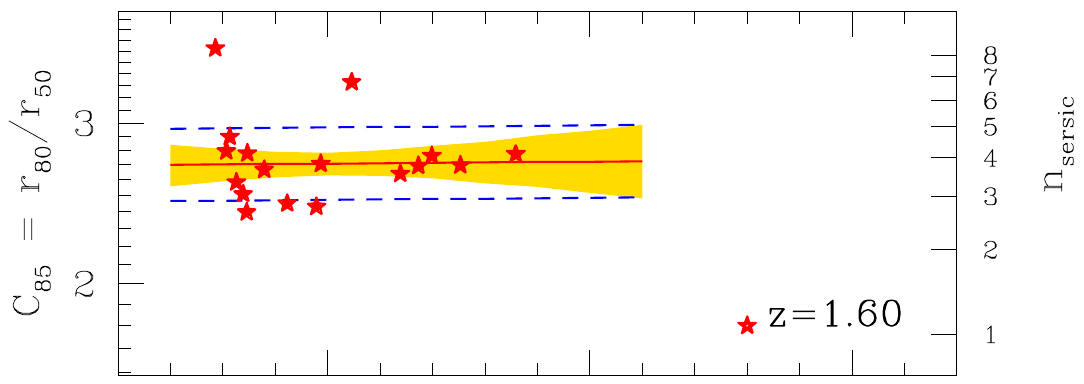}}
\centerline{\includegraphics[width=9truecm]{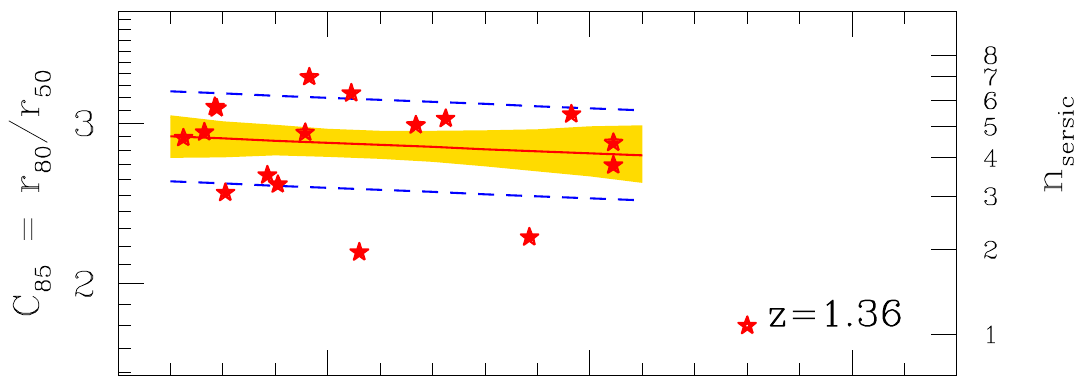}\includegraphics[width=9truecm]{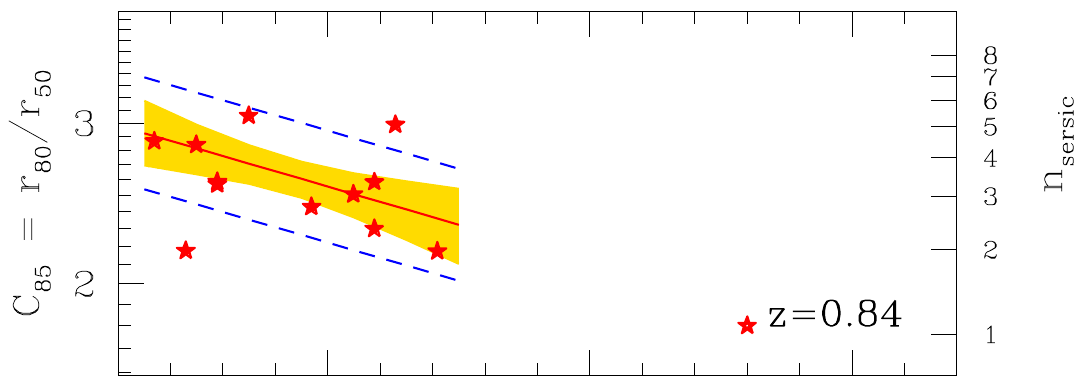}}
\centerline{\includegraphics[width=9truecm]{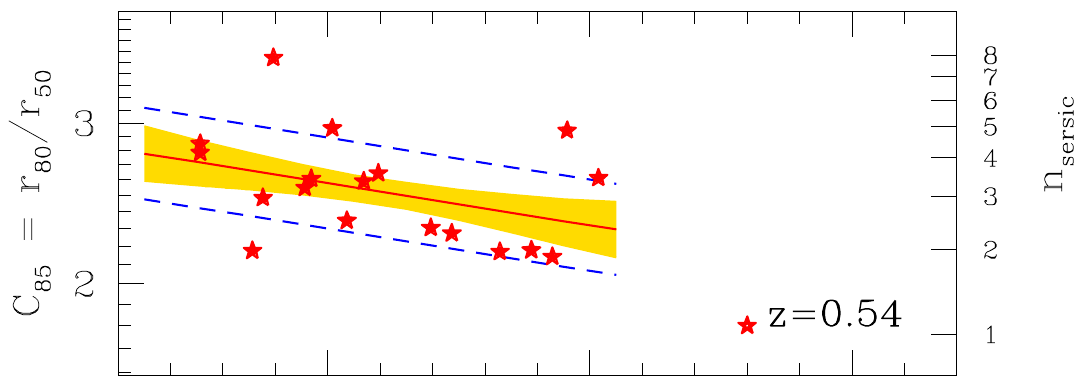}\includegraphics[width=9truecm]{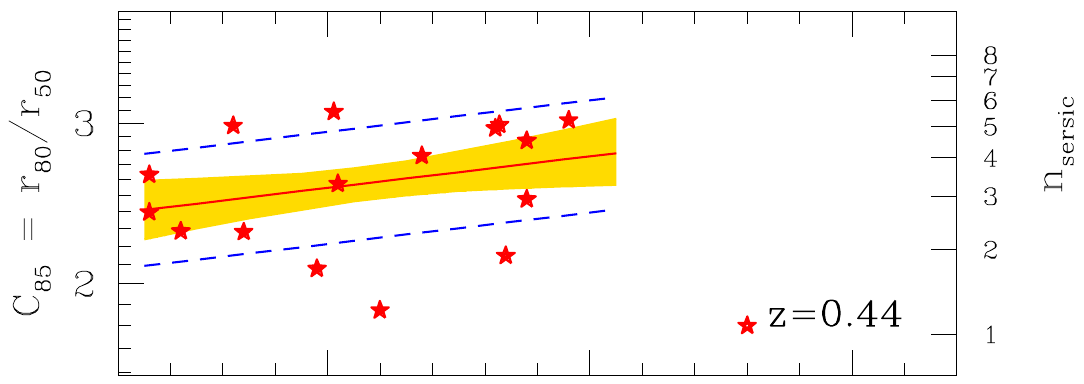}}
\centerline{\includegraphics[width=9truecm]{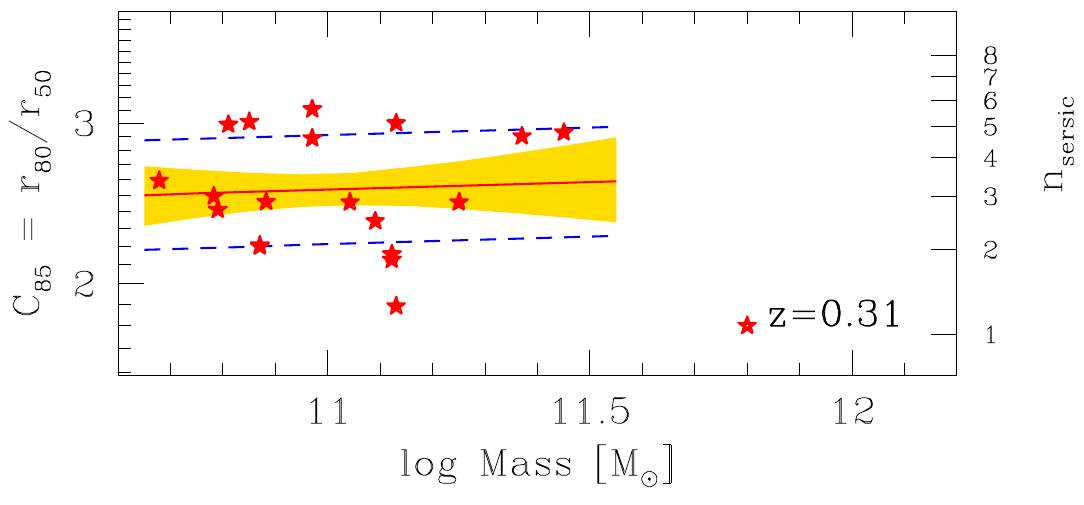}\includegraphics[width=9truecm]{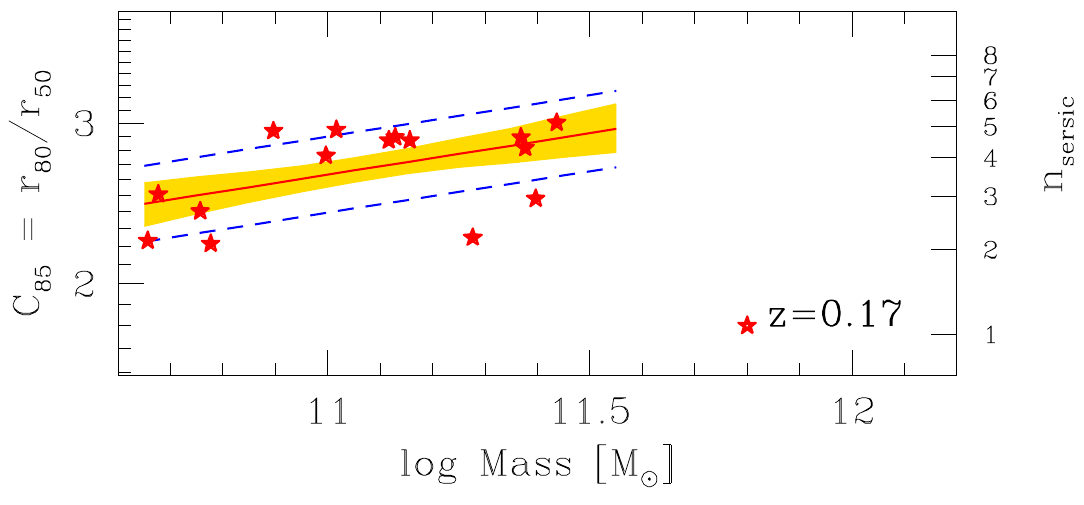}}
\caption[h]{Concentration-mass relation of red-sequence early-type cluster galaxies.
The yellow shading indicates the 68\% uncertainty (posterior higher density), the dashed 
blue corridor indicates a$\pm 1\sigma$
range around the mean model.
Concentrations are corrected for (negligible) PSF blurring effects.
}
\end{figure*}

\begin{figure}
\centerline{\includegraphics[width=9truecm]{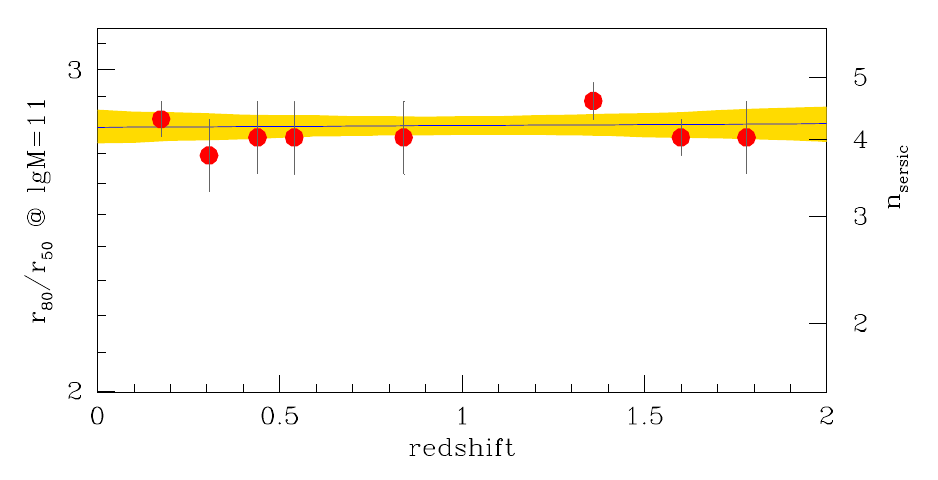}}
\caption[h]{Concentration at $\log M/M_\odot=11$ vs redshift. 
The solid line and shading show the fitted 
relation and its 68\% uncertainty (posterior highest-density interval).
}
\end{figure}

\begin{figure}
\centerline{\includegraphics[width=9truecm]{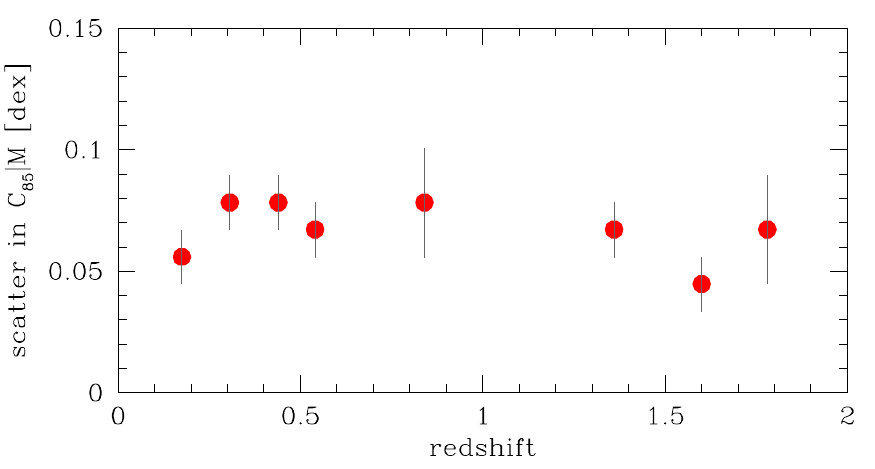}}
\caption[h]{Scatter around the concentration-mass relation vs redshift. 
Scatter is computed from Student-t scale $s$ as $\sigma = s *\sqrt{10/8}$.
}
\end{figure}

\begin{figure}
\centerline{\includegraphics[width=9truecm]{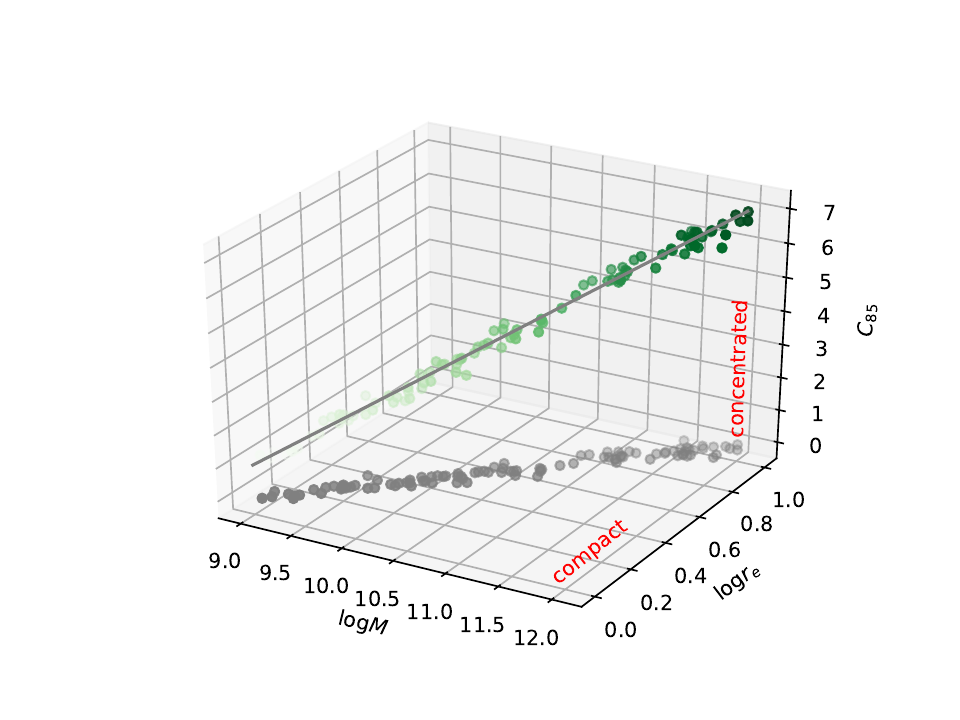}}
\caption[h]{Sketch view, with arbitrary points, of the mass-size-concentration 
space occupied by galaxies. 
}
\end{figure}

\section{Summary and discussion}

\subsection{Lexical preamble}

As emphasized in the introduction and sketched in Fig.~8, a) concentration is a measure 
orthogonal to size and b) compactness is  a measure of the form of the profile;
exponential profiles have smaller values of the index $C_{85}$
than de Vaucouleurs profiles, no matter
the size, compactness, or mass of the two compared profiles. 
The two terms should not be interchanged; for example, high redshift galaxies are 
not more concentrated because they are smaller 
(when, for example, they have the same Sersic or $C_{85}$ index).
Furthermore, the concept of compactness assumes the existence of a spread in size
at a fixed mass;
a system is not compact because its effective radius is small, but because there
are other larger galaxies, that is to say there is a spread in size\footnote{No car can be called
compact if all cars have one single size.}. For this reason we
modeled the spread vertically in the mass-size plane, instead of, for instance, 
studying the (dis)appearance of compact 
objects by defining compactness using an absolute size, or density,
threshold. The latter choice mixes the target measurement with 
a variation in the scatter at a given mass and with the evolution of the mean relation\footnote{For
example, to broadly categorize a person as underweight, 
the body mass index is used, which is tied to height (size) to account for the change of 
height with age.}.
Finally, 
measurements performed on observational data
are made at a fixed mass (or a fixed number density, or at the fixed quantity that is used
for measurements), and this should be kept in mind when interpreting 
the results.
We can consider, for example, the universe simulated in Naab et al. (2009)
formed by early-type galaxies 
that grow in size by a factor of 2.6 while their mass growth is 0.3 dex between $z=2$ to $z=0$. 
We can assume
that the mass-size relation has a slope of $\sim 0.5$, in agreement with observations.
Some studies would have concluded that galaxies were 0.4 dex ($=\log 2.6 $) smaller at high redshift. Other studies
would have instead reported that galaxies are 0.25 dex smaller 
at high redshift
at a fixed mass. In fact, galaxies that are 0.3 dex more massive have 0.15 dex larger
sizes because of the mass-size relation, and therefore the
growth at a fixed mass is $0.25$ dex. 
In short, evolution is partially along the mass-size relation. 
Some studies would have  
suggested that in the above universe galaxies were compact at $z=2$, but
since there is no spread in size at a fixed mass in that universe, 
no galaxy is compact whatever its redshift may be. At most, $z=2$ galaxies might be called 
small compared to present-day standards.

To summarize, the orthogonal concepts of compactness and concentration must not be confused; compactness deals with size and needs a spread at fixed mass, 
while concentration deals with the curvature of the surface brightness profile. 
Our concentration index
is sensitive to the curvature outside the 
effective radius by the way it is defined and because of the $\sim 1$ kpc 
resolution of the  data at high redshift. 
Instead, a hypothetical concentration index involving the 20\% flux radius would require the measurement of
fluxes in apertures smaller than the WFC3 pixel size at all masses at high redshift.

\begin{figure}
\centerline{\includegraphics[width=6truecm]{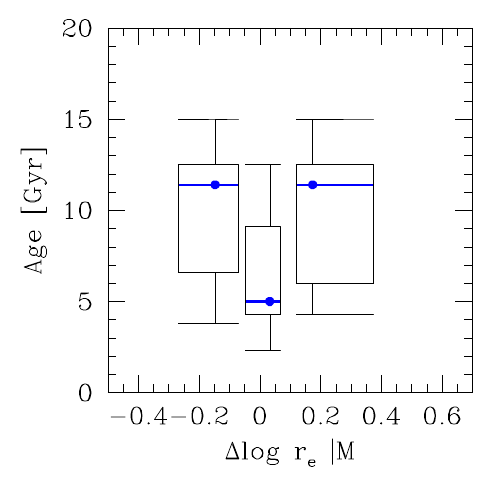}}
\caption[h]{Age distribution of Coma early-type galaxies 
small, average, and large for their mass, respectively. 
The plot is a standard box-whisker; the vertical box width
delimits the 1$^{st}$ and 3$^{rd}$ quartile, while the median (2$^{rd}$ quartile)
is indicated by the blue horizontal  segment inside the box.
The horizontal box width gives the full x range of each bin,
while the error bars reach the minimum and maximum in the ordinate. 
Observed ages can be older than the Universe age because
of errors. 
}
\end{figure}

\subsection{Summary of main results}

Using an almost random sampling of a mass-limited sample formed by 128 morphologically early-type galaxies
in clusters with $\log M/M_{\odot} \ga 10.7$ spanning the wide range $0.17<z<1.81,$ 
 we find that in the last 10 Gyr 
both half-light and 80\% light sizes increase
by a factor of $1.7$, and that concentration stays constant within $2$\%, that is to say the size
growth of early-type galaxies in cluster environments proceeds at the same pace at both 
radii.
The scatter around
the mean relation measures the amount of variability, from system to system, 
of the amount of dissipation that leads to the observed galaxy. 
The constancy of the scatter around the size-mass relation with redshift  (Fig.~4)
implies that dissipation does not vary greatly with epoch.

\subsection{Morphological sample selection is key}

Unlike most published mass-size relations, our sample is morphologically
selected. Only one-third of passive galaxies have an early-type morphology (Paper I).
Indeed, passive galaxies are formed by a mix of early-type galaxies, 
recently quenched galaxies, and dusty star-forming galaxies (Williams et al. 2009,
Moresco et al. 2013, Carollo et al. 2013, Paper I, Andreon 2018) and splitting this
composite sample into more homogeneous parts is key to discriminate
size or concentration evolution from a change of the morphological mix. In fact, 
morphological classification allowed
De Propris et al. (2016) to properly interpret the measured apparent
concentration change.

In literature,
the Sersic index is often used to select
the sample of early-type galaxies (e.g., Newman et al. 2014; De Propris et al. 2016; 
Strazzullo et al. 2010 for clusters;  for galaxies in the field this is the rule). 
Our sample is morphologically selected and as such avoids the complications
of studying the evolution in concentration of a concentration-selected sample;
if galaxies of a given morphological type
with a given value of concentration are missing, it is because 
Nature does not produce them.

\subsection{Size growth: intrinsic vs entirely due to a mass growth}

Since we find a size growth at fixed mass, and size itself depends on mass, 
a natural question is to what extent the measured size growth is just mirroring 
a possible mass growth.
Between the $z=2$ to $z=0$, stellar mass is barely growing, if at all. Mass and luminosity function determinations of early-type galaxies in cluster  
concur to find a non-evolving massive end, with upper limits of $\sim 0.1$ dex 
change between
$z=2$ and $z=0$ (Andreon 2006, 2012, Andreon et al. 2014 and references therein), given
some assumptions,\ such as which
synthesis population model and age are used (see for example Andreon et al. 2014).
A neglected $0.1$ dex mass change would spuriously bias the evolution at fixed mass 
by $0.05$ dex in size because 
of the $\sim 0.5$ slope
of the mass-size relation. However,
the observed evolution at a fixed mass is $0.22$ dex, well beyond the effects of
a hypothetically neglected 0.1 dex evolution in mass.
To observe a spurious $0.22$ dex evolution at fixed mass,
we would need a mass evolution four times larger than what is allowed by observations.

\subsection{The lack of scenario explaining all the observations}

While both the mean $r_{80}$ or $r_{50}$ sizes change by a factor of 1.7 (see Sect.~3.1),
concentration stays constant (within 2\%, Sect.~3.2). The spread in concentration is
small ($\sim 15$\%) and constant (Fig.~7). 
Now we consider some unsuccessful attempts to find
a physical mechanism producing trends in agreement with the observations. 

First, progenitor bias has been
already discussed and discarded in Paper I. Figure~9 reiterates the point,
this time using ages of Coma galaxies from Harrison et al. (2010), chosen because they were used
by Saracco et al. (2020) to claim the existence of a progenitor bias. Galaxies
below, on, and above the mass-size relations have much the same median age and
distribution of ages. To induce a bias in the intercept of the
mass-size, there should instead be a trend in age as a function of size residuals
at a fixed mass (abscissa), not visible in Fig.~9, 
independently confirming the similar result in Paper I using
Smith et al. (2012) ages. 

Second, truncation by the cluster halo should, at first glance,  affect $r_{80}$  more than
the inner radius $r_{50}$, and therefore should decrease the value of the $C_{85}$ index. Given that we
instead observed increasing sizes 
and constant values of $C_{85}$, truncation seems to be
excluded by the data. Furthermore,  baryons
are distributed in a more compact way than dark matter in galaxy halos (Limousin
et al. 2009) and therefore truncation
seems to be excluded a priori, as well as by observations.

Third, mergers, either major or minor, seems to be excluded because major mergers 
mostly increase 
mass, whereas minor mergers mainly change concentration, at least in
simple settings. Idealized simulations of galaxies 
living outside a massive halo (cluster) in a non-cosmological setting  predict
an inside-out growth accompanied by a mass growth (Nipoti et al. 2003; 
Naab et al. 2009; Hilz et al. 2013), with amplitudes depending on 
which publication is considered and on whether the accretion is minor or major, 
unlike what we observe for galaxies. 
Moving to cosmological simulations,
our  negligible evolution of concentration with cosmic time of early-type galaxies in cluster is 
instead in line with the results of numerical simulation of massive galaxies 
in Tacchella et al. (2019). However,
size and mass evolution of simulated galaxies grossly mismatch the observed evolution,
perhaps due to differences in environments between simulated and observed galaxies.
Overall, it is unclear whether results obtained on
simulations of galaxies outside large massive halos, either idealized or cosmological,  
can be applied for galaxies inside a large massive halo.
Nevertheless, Tacchella et al. (2019) simulations show that it is possible to change galaxy sizes
without changing the galaxy concentration, though still changing the galaxy
mass. 

A meaningful comparison that allows us to shed light on which mechanisms are operating
in galaxy clusters would require cosmological simulations of galaxies in massive
clusters.
However, current simulations
either do not have the $<1$ kpc resolution for measuring galaxy
size and concentration, or do not sample volumes large enough to include rich 
galaxy clusters.\ There is the notable exceptions of a few cluster re-simulations, such as
C-EAGLE  (Barnes et al. 2017), which, however, do not have any published predictions.

\subsection{A potential bias on estimate of structure evolution}

With rare exceptions, the half-light, or half-mass, radius is identified with the
galaxy size in literature, and its evolution is interpreted as the
evolution of the whole galaxy structure.    
The inferred evolution of galaxy structure, commonly derived using $r_{50}$ sizes,
seems not to be biased by the arbitrary choice of 
using half-light radii, at least for the radii 
considered in this work and for early-type galaxies
in cluster.

\subsection{Comparison with previous works}

Compared to color gradient measurements (e.g., 
De Propris et al. 2015, Chan et al. 2016, 
Ciocca et al. 2017), our study, measuring the relative rate at which the main body and the outskirts are built,
has the advantage of directly assessing the relative structure build-up. 
In fact, under the optimistic assumption that color gradients are only
sensitive to age differences,
a given 
color gradient may
imply a 
huge mass accretion or merging, or not at all, 
depending on the unknown age difference between accreted 
and already-present stellar
populations. For a similar reason, spatially unresolved spectroscopy 
(e.g. Saracco et al. 2020, Stockman et al. 2020, Matharu et al. 2020) 
is inadequately informative about the
relative build-up of the central and outer parts of the galaxies; spatially-unresolved
spectra measure stellar ages, averaged over a large portion of the galaxy, 
not assembly time of the galaxy parts.

To our best knowledge, there are no directly comparable studies on the
evolution of concentration of morphological early-type 
galaxies in clusters. Our study, measuring the relative growth of the galaxy
main body and outskirts, goes beyond
recent works on the mass-size relations of galaxies in clusters
(Kelkar et al. 2015, Kuchner et al. 2017, Sweet et al. 2017, 
Morishita et al. 2017, Saracco et al. 2017, Matharu et al. 2019). Lacking 
published measurements of the  
$r_{80}$-mass relation, of the concentration-mass relation, 
or of the concentration evolution, we cannot compare
relations derived by different authors
or galaxy populations in clusters.

Even allowing the environment to be different,
the literature is quite scarce. 
A precise comparison of the concentration evolution of
related, yet different, classes of galaxies residing in different environments,
such as Gu et al. (2019) vs our study, is extremely difficult; it is difficult to constrain a problem with three covariates (population, environment, and amplitude
of the progenitor bias) with two measurements.
A meaningful comparison is easier at a fixed environment. Such a comparison is  presented in a companion paper measuring concentration evolution in the field (Andreon, in preparation).

\section{Conclusions}

We measured the radii, including 80\% of the light, of 
an almost random sampling of a mass-limited sample formed by
128 morphologically early-type galaxies
in clusters with $\log M/M_{\odot} \ga 10.7$ spanning the wide range $0.17<z<1.81$. We measured concentration
by combining it with the half-light radius derived in Paper I. Because of the adopted
non-parametric estimate of concentration, it
focuses on the outer region 
of the galaxy and is insensitive to the profile inside the effective radius, unlike the
Sersic index.

We found that concentration stays constant within $2$\% while both 50\% and 80\% light radii change
by a factor of $1.7$ from $z=2$ to $z=0$ at a fixed mass. We ruled out, for the second
time, progenitor bias as a spurious source of size growth using an independent 
set of ages. We also ruled out mass growth as an
entire source of size growth; to explain a $0.22$ dex evolution in size,
a $0.44$ dex growth in mass is needed, while only at most $0.1$ dex is allowed
by mass function measurements.

Existing physical explanations proposed in the literature
are unable to consistently explain the changing
radii at a fixed mass while keeping concentration constant when mass evolution is
negligible, which
call for adopting or performing cluster simulations in a cosmological setting with a sufficient
resolution (better than 1 kpc, to resolve the effective radius) to identify the
physical mechanism affecting the galaxy structure.

\begin{acknowledgements}
SA thanks the anonymous referee for his/her valuable report,
S. Tacchella for providing their concentrations updated for
the $C_{85}$ index, C. Nipoti for useful comments, P. Saracco for useful discussions, and 
C. Bernasconi, B. Garilli, C. Giorgieri, 
and M. Marelli for efforts in 
recovering data saved 25 years ago in a nowadays obsolete device. 
It's a pleasure to thank the late Raymond Michard, whose original code turned
out to be easy to modify to compute the concentrations used in this work.
\end{acknowledgements}

{}

\end{document}